\documentclass{article}

\usepackage{arxiv}

\usepackage[utf8]{inputenc} 
\usepackage[T1]{fontenc}    
\usepackage{hyperref}       
\usepackage{url}            
\usepackage{booktabs}       
\usepackage{amsfonts}       
\usepackage{nicefrac}       
\usepackage{microtype}      
\usepackage{lipsum}		
\usepackage{graphicx}
\usepackage{multirow}
\usepackage{cite}

\usepackage{amsmath}
\usepackage{chemformula}

\usepackage{epstopdf}
\usepackage{array}
\usepackage{subfigure}
\usepackage{ragged2e}
\usepackage{booktabs,makecell,multirow,tabularx}
\usepackage[ruled,linesnumbered]{algorithm2e}

\title{PaRO-DeepONet: a particle-informed reduced-order deep operator network for Poisson solver in PIC simulations}

\author{ jianhua Lv, ~Linlin Zhong\thanks{This paper is currently under consideration by a journal.}  \\
	School of Electrical Engineering\\
	Southeast University\\
	No.2 Sipailou, Nanjing, Jiangsu Province 210096, P. R. China\\
	\texttt{linlin@seu.edu.cn}\\
}

\date{April 25, 2025}

\renewcommand{\headeright}{}
\renewcommand{\undertitle}{A Preprint}

\begin{document}
\maketitle

\begin{abstract}
Particle-in-Cell (PIC) simulations are widely used for modeling plasma kinetics by tracking discrete particle dynamics. However, their computational cost remains prohibitively high, due to the need to simulate large numbers of particles to mitigate statistical noise and the inefficiency in handling complex geometries. To address these challenges, we propose PaRO-DeepONet, a particle-informed reduced-order surrogate framework that integrates Proper Orthogonal Decomposition (POD) with Deep Operator Network (DeepONet). By performing manifold sampling of particle evolution in PIC simulations and constructing snapshot matrices from various particle states, PaRO-DeepONet applies POD projection to extract latent charge density features that are shared across similar plasma scenarios. These reduced features are fed into the branch network of DeepONet, enabling efficient mapping from charge distributions to electrostatic potential fields. We validate PaRO-DeepONet on four representative benchmark cases. The model consistently achieves relative L$^2$ errors below 3.5$ \% $, with the best case reaching as low as 0.82$ \% $. Notably, although the charge deposition becomes highly discrete due to particle sparsity, PaRO-DeepONet still reconstructs smooth and continuous potential fields, demonstrating strong robustness to sparse inputs and excellent generalization beyond traditional grid-based solvers. In terms of computational performance, PaRO-DeepONet reduces total PIC simulation time by 68$ \% $–85$ \% $ and cuts the Poisson solver runtime by up to 99.6$ \% $. These results highlight the potential of PaRO-DeepONet as a robust and scalable solution for accelerating PIC simulations, especially in complex or high-dimensional geometries. By significantly reducing computational cost while maintaining high accuracy, this work establishes machine learning–based operator surrogates as a promising new paradigm for next-generation kinetic plasma simulations.
\end{abstract}

\section{INTRODUCTION}
\label{sec:sec:sec1}
\paragraph{}
The Particle-in-Cell (PIC) method is a widely adopted and powerful computational approach for plasma modeling. Grounded in first principles, it effectively simulates plasma behavior under a wide range of conditions and has found extensive applications in fields such as semiconductor etching \cite{1,2,3}, fusion research \cite{4,5,6}, and laser–plasma interactions \cite{7,8,9}. Given the transformative significance of these applications, enhancing both the accuracy and efficiency of PIC simulations is of paramount importance.
\paragraph{}
In PIC simulations, a finite number of macroparticles—typically ranging from thousands to hundreds of thousands—are sampled from the full distribution function and mapped onto a spatial grid. The electrostatic interactions among particles are then resolved by solving partial differential equations (PDEs), typically via the finite difference method (FDM), which often becomes the most time-consuming step. Even with multigrid acceleration techniques \cite{10}, solving a single initial plasma condition can take several days or even weeks. Moreover, particle sampling and grid discretization introduce artificial noise, affecting the overall solution accuracy \cite{11,12,13}. Therefore, improving the efficiency and accuracy of PDE solvers within PIC frameworks would significantly advance the state of plasma modeling.
\paragraph{}
Recently, deep learning has emerged as a promising alternative for solving PDEs. According to the universal approximation theorem \cite{14}, neural networks with even a single hidden layer can approximate any nonlinear continuous function, offering both generalization and interpretability. Among deep learning-based PDE solvers, two prominent frameworks are Physics-Informed Neural Networks (PINNs) \cite{15} and Deep Operator Network (DeepONet) \cite{16}. PINNs incorporate governing physical laws directly into the loss function, eliminating the need for grid discretization and offering flexibility for complex geometries. Notably, PINNs have been successfully applied to solve the Boltzmann equation and thermal plasma models, significantly accelerating fluid-based plasma simulations \cite{17,18,19}.
\paragraph{}
In contrast, DeepONet act as powerful operator learners, mapping input functions to output functions across high-dimensional spaces using spatiotemporal coordinates. DeepONet have demonstrated strong generalization capabilities and perform well across various grid resolutions \cite{16}. Owing to their intrinsic spatiotemporal properties, these models show potential for mitigating the noise introduced by particle discretization. Despite their promise, applications of DeepONets in plasma physics remain limited, likely due to the inherent complexity of coupling multiple physical fields in plasma systems. Peng et al. \cite{20} combined DeepONet with PINNs to solve Poisson and convection-diffusion equations in streamer discharge, achieving improved efficiency and accuracy. Wang et al. \cite{21} proposed the Deep Cross Section Network (DeepCSNet) to predict differential ionization cross-sections, with excellent generalization to unknow molecules. Gopakumar et al. \cite{22} utilized Fourier Neural Operators (FNO) to construct a surrogate model for tokamak fusion plasmas, enabling accurate spatiotemporal predictions from diagnostic camera data.
\paragraph{}
One of the key challenges in applying neural networks to PIC simulations lies in efficiently encoding high-dimensional discrete particle information without substantial loss of physical fidelity. In fluid dynamics, reduced-order models (ROMs) have been developed to represent high-dimensional systems using low-dimensional latent structures \cite{23}. Among these, Proper Orthogonal Decomposition (POD) is a classical and efficient method that has been successfully combined with machine learning to reduce computational costs \cite{24,25,26}. For example, Heredia et al. \cite{24} used POD with deep learning to rapidly predict fluid evolution, while Ahmed et al. \cite{25} integrated POD with autoencoders to enhance accuracy and reduce training complexity. Although particle-based models differ fundamentally from fluid models, these approaches inspire the possibility of compressing particle data for integration into deep networks.
\paragraph{}
In this work, we propose a novel reduced-order surrogate model for the Poisson solver in PIC simulations by incorporating particle-informed ROM techniques into the DeepONet framework, referred to as PaRO-DeepONet. The model effectively integrates particle data into a neural operator architecture and demonstrates feasibility and generalization capability across four case studies involving full particle evolution. Owing to the strong performance of DeepONet, the proposed framework offers a promising, efficient, and extensible solution for accelerating PIC simulations, particularly in more complex plasma systems.
\paragraph{}
The rest of the paper is organized as follows: Section II introduces the proposed PaRO-DeepONet in detail, including the manifold sampling strategy, the implementation of particle-informed POD for latent feature extraction, and its integration as a PDEs solver within the PIC framework. Section III presents four numerical case studies: 2D/3D electrostatic sphere diffusion, electron diffusion in an L-shaped domain, and a case analogous to electron extraction in neutral beam injectors. Finally, conclusions are drawn in Section IV.

\section{METHOD}
\paragraph{}
In a typical electrostatic Particle-in-Cell (PIC) simulation, the charge deposition algorithm is first employed to obtain the charge density distribution $ \rho $(x, y, z, t) on the spatial grid. Subsequently, the potential distribution  $ \varphi $(x, y, z, t) is computed by solving the Poisson equation. While the mapping from  $ \rho $ to  $ \varphi $ may appear straightforward, and several operator learning frameworks (e.g., DeepONet \cite{16}, FNO \cite{27}, POD-DeepONet \cite{26}) have demonstrated strong performance on similar problems, applying these models in the context of PIC simulations introduces unique challenges.
\paragraph{}
Specifically, the computed potential at each time step is used to derive the electric field E, which in turn governs the motion of particles. This dynamic coupling implies that even minor errors in the potential solution can accumulate over time, ultimately resulting in significant deviations in particle trajectories. Consequently, compared to generic PDE-solving tasks, the integration of operator networks as surrogate PDE solvers in PIC simulations requires exceptionally high accuracy and robust global feature extraction capabilities.
\paragraph{}
This section introduces the core algorithm of PaRO-DeepONet, a particle-informed reduced-order surrogate model designed to replace traditional Poisson solver in PIC frameworks. The proposed method integrates a reduced-order representation of the charge density distribution via POD, combined with DeepONet to directly approximate the mapping from  $ \rho $ to  $ \varphi $. By compressing the high-dimensional particle data into a low-dimensional latent space and learning the corresponding operator with DeepONet, PaRO-DeepONet significantly accelerates the solution of PDEs while maintaining high accuracy throughout time-evolving PIC simulations.

\label{sec:sec:sec2}
\subsection{Particle-informed reduced-order surrogate framework}
\label{sec:sec2.1}
\begin{figure}
	\centering
	\includegraphics[width=9.5cm]{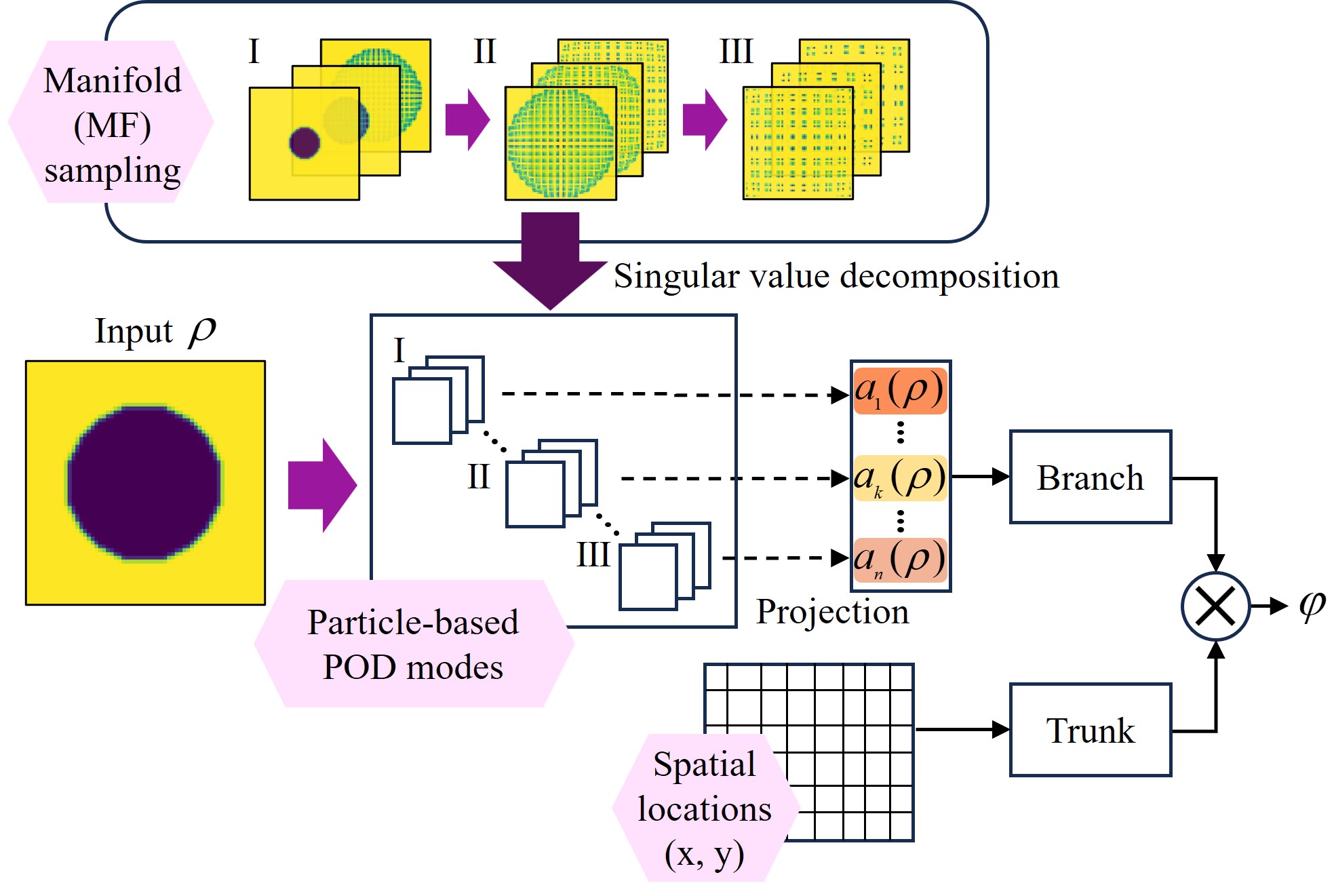}
	\caption{Schematic of the PaRO-DeepONet framework integrating manifold sampling, particle-informed POD, and DeepONet.}
	\label{fig:fig1}
\end{figure}

\paragraph{}
The proposed PaRO-DeepONet framework consists of three key components: manifold sampling, particle-informed Proper Orthogonal Decomposition (POD), and the Deep Operator Network (DeepONet), as illustrated in Figure \ref{fig:fig1}. Manifold sampling is used to construct the snapshot matrix from the time-evolving charge density fields in the PIC simulations. POD then extracts the dominant modes from this manifold, effectively capturing the key patterns in particle-induced charge dynamics. This process compresses the high-dimensional charge distribution into a low-dimensional latent representation, making it amenable to operator learning via DeepONet. Once the reduced-order charge density and corresponding spatial coordinates are obtained, DeepONet is employed to learn the operator mapping from charge density to electrostatic potential.

\paragraph{}
As the core component of this framework, DeepONet was originally proposed by Lu et al. \cite{16}. It is an operator learning architecture that approximates nonlinear mappings between infinite-dimensional function spaces. DeepONet consists of two subnetworks: the branch network, which encodes the input function $u$ evaluated at fixed sensor locations, and the trunk network, which encodes the target spatial-temporal coordinates $ \zeta $. The final output $ G $ of the learned operator $ \widehat{G}  $  is computed as the dot product between the outputs of these branch and trunk networks:

\begin{equation}
\label{equ:equ2}
\widehat{G} \left( u \right)\left( \zeta  \right) = \sum\limits_{k = 1}^p {{b_k}} \left( {{u_1},{u_2},...,{u_m}} \right){t_k}\left( \zeta  \right)
\end{equation}
where $ b_k $ and $ t_k $ are the outputs of branch and trunk networks, respectively.

\paragraph{}
To enhance spatial representation efficiency, several variants of DeepONet have been proposed. One notable extension is POD-DeepONet \cite{26}, which utilizes POD basis functions as inputs to the trunk network. While effective, this framework is designed for general-purpose operator learning and does not account for the specific challenges posed by PIC simulations, such as discrete particle sampling and dynamic feedback between fields and particles. The PaRO-DeepONet architecture proposed in this study differs from POD-DeepONet and incorporates adaptations tailored to PIC simulations. Detailed explanations of the architectural modifications and implementation strategies are discussed in Section 2.1.3, and the case-specific configurations are presented in Section 3.

\subsubsection{Manifold sampling}
\label{sec:sec:sec2.1.1}
\paragraph{}
In electrostatic PIC simulations, under similar initial conditions—such as the number of macroparticles, their spatial distribution, and boundary conditions—the trajectories of charged particles exhibit coherent evolution governed by self-consistent fields. As a result, the system dynamics can be approximated as evolving on a low-dimensional manifold embedded within the high-dimensional state space \cite{28}.

\paragraph{}
To effectively sample this underlying manifold \cite{29}, we perform a set of $m$ independent PIC simulations, each initialized with distinct but physically plausible initial conditions. These simulations are allowed to proceed until the number of particles remaining in the computational domain drops below 10$ \% $ of the initial count. Beyond this point, the spatial distribution of the charge density becomes increasingly discontinuous, and therefore unsuitable for representing a smooth manifold. In systems with highly sparse particle distributions or sensitive dynamics, a higher termination threshold may be necessary.

\paragraph{}
During each simulation, the charge density distribution is sampled at uniform time intervals based on the PIC time step. This process yields a time series of charge density fields for each initial condition, capturing the spatiotemporal evolution of the system. The ensemble of sampled trajectories forms a quasi-linear approximation of the charge density manifold, and is formally represented as:

\begin{equation}
\label{equ:equ2}
{D_{{\rm{man}}}}\left\{ {{\rho _1}\left( t \right),{\rho _2}\left( t \right),...,{\rho _{\rm{m}}}\left( t \right)} \right\}
\end{equation}

Where $ \rho_i(t) $ denotes the sequence of charge density distribution over time in the $i$-th simulation trajectory. This manifold sampling dataset serves as the foundation for the POD-based dimensionality reduction, enabling low-dimensional yet physically meaningful representation of the complex charge dynamics for subsequent operator learning.

\subsubsection{Particle-informed POD}
\label{sec:sec:sec2.1.2}
\paragraph{}
To reduce the dimensionality of the charge density distribution sampled from $D_{{\mathrm{man} }} $, we employ Proper Orthogonal Decomposition (POD), a widely used technique for extracting dominant spatial modes from high-dimensional data. Let the number of grid points in the simulation domain be $n$, and the number of time steps for each simulation be $ t\mathrm{max} $. Then, for each manifold sample ${\rho _i}(t) \in {D_{{\rm{man}}}}$, we flatten each 3D charge density distribution at each time $t$ into a 1D column vector of size $n$. By stacking these vectors over all time steps and all simulations, we obtain a global snapshot matrix of size $n\times(m\cdot t\mathrm{max} )$, which captures the full range of spatiotemporal charge dynamics.

\paragraph{}
However, the spatial distribution of the charge density is strongly influenced by the number of particles remaining in the domain. As particles exit the computational domain, the deposited charge becomes increasingly sparse and discrete, thereby reducing the statistical consistency across snapshots. To account for this, we partition the snapshot matrix into 10 subsets $\left\{ {{A_0},{A_1},...,{A_9}} \right\}$ based on the ratio $r = \frac{{{N_{{\rm{particles}}}}\left( t \right)}}{{{N_{{\rm{initial}}}}}}$, where$N_{\mathrm{paritcles}}(t)$ is the number of particles remaining at time $t$, and $N_{\mathrm{initial}}$ is the initial particle count. Each subset $A_k$ includes snapshots for which this ratio lies within the interval$[1-0.1(k+1),1- 0.1k)$. This partitioning ensures that snapshots in each subset share similar particle statistics, preserving the consistency required for effective mode extraction.

\paragraph{}
For each subset matrix ${A_k} \in {R^{n \times {T_k}}}$, where $T_k$ is the number of snapshots in the $k$-th group, we perform Singular Value Decomposition (SVD) \cite{30}:

\begin{equation}
\label{equ:equ3}
{A_k} = {U_k}{\sum _k}V_k^* + \overline {{A_k}} 
\end{equation}

Here, $\overline {{A_k}}$ is the row-wise mean vector of $A_k$, ${U_k} = [\phi _1^{\left( k \right)},\phi _2^{\left( k \right)},...] \in {R^{n \times n}}$ contains the spatial POD modes (left singular vectors), $ \Sigma_k $ is the diagonal matrix of singular values arranged in descending order, and $V_k$ contains the temporal coefficients.

\paragraph{}
To retain only the most informative spatial modes, we apply energy-based truncation using the Relative Information Criterion (RIC):

\begin{equation}
\label{equ:equ4}
{\rm{RIC}}(\% ) = \frac{{\sum\limits_{i = 1}^r {\sigma _i^2} }}{{\sum\limits_{i = 1}^m {\sigma _i^2} }} \times 100
\end{equation}

where RIC is the relative information criterion, $r$ is the minimum order to achieve the required energy, and ${\sigma_i}$ are the singular values of $\Sigma_k$. We select the smallest integer $r_i$ such that RIC($r_i$) $\ge$  $\eta$, where $\eta$ is a user-defined threshold (e.g., 0.999) indicating the proportion of energy to retain. This yields a reduced set of $r_i$ modes for each $A_k$.

\paragraph{}
The high-dimensional charge density distribution $\rho (t) \in R$ is then projected onto the retained spatial basis to obtain a low-dimensional coefficient vector:

\begin{equation}
	\label{equ:equ5}
	\rho (t) = \sum\limits_{k = 1}^r {{a_k}} \left( t \right){\phi _k} + \overline {{A_k}} 
\end{equation}

\paragraph{}
These POD coefficient vectors provide a compact information-preserving representation of the charge density field and serve as the input to the downstream DeepONet model.

\subsubsection{PaRO-DeepONet}
\label{sec:sec:sec2.1.3}

\paragraph{}
Following the dimensionality reduction of the charge density distribution via POD, each snapshot is projected onto its corresponding spatial modal basis $U_k$, yielding the mode coefficient vector $ a_k = [a_1, a_2, …, a_{rk}]$, where $ k\in [0, 9] $ denotes the particle state interval as defined in Section 2.1.2. 

\paragraph{}
To effectively capture the spatiotemporal dynamics of charge evolution in the grids, PaRO-DeepONet extends beyond the use of a single mode vector $a_k$. Instead, it concatenates multiple modal vectors corresponding to recent particle states to form a unified input representation: $a' = [a_0; a_1; …; a_k] $, which serves as the input function to the branch network. This aggregation enables the network to incorporate temporal context from prior particle states. Such historical information is critical for accurately modeling the time-dependent behavior of charge evolution in electrostatic PIC simulations.

\paragraph{}
Simultaneously, the trunk network receives the spatial coordinate $\zeta $ of the target grid point, defined as $\zeta  = \left[ {x;y} \right]$ in 2D or $\zeta  = \left[ {x;y;z} \right]$ in 3D. In this work, both the branch and trunk networks are implemented as fully connected feedforward neural networks, with the number of layers and neurons configurable based on the complexity of the charge density distribution.

\paragraph{}
The PaRO-DeepONet outputs the predicted potential $\widehat{\varphi }\left( t \right)$  at each location and time $t$ via the standard DeepONet formulation:

\begin{equation}
	\label{equ:equ6}
\widehat{\varphi }\left( t \right) = Branch\left( {a'\left( {\rho \left( t \right)} \right)} \right) \cdot Trunk\left( \zeta  \right)
\end{equation}

where $\rho \left( t \right)$ is the charge density distribution, $a'$ is the modal coefficient vector, $\zeta$ is the grid coordinate, and  $Branch\left ( \cdot  \right ) $  and $Trunk\left ( \cdot  \right )$ are the corresponding network outputs. The loss function of PaRO-DeepONet is expressed as follows: 

\begin{equation}
	\label{equ:equ7}
L = \frac{{{{\left\| {\mathord{\buildrel{\lower3pt\hbox{$\scriptscriptstyle\frown$}} 
						\over \varphi }  - \varphi } \right\|}_2}}}{{{{\left\| \varphi  \right\|}_2}}}
\end{equation}

where  $\widehat{\varphi } $ and $\varphi$  are the predicted and ground-truth potentials, respectively, and the summation should be run over all training points.

\paragraph{}
To handle complex particle boundaries commonly encountered in PIC simulations, the performance of PaRO-DeepONet can be further improved via hard boundary encoding or regularization strategies. Further implementation details are provided in Sections 3.3 and 3.4.

\label{sec:sec2.2}
\subsection{Electrostatic PIC simulation based on PaRO-DeepONet}
\begin{figure}
	\centering
	\includegraphics[width=9.5cm]{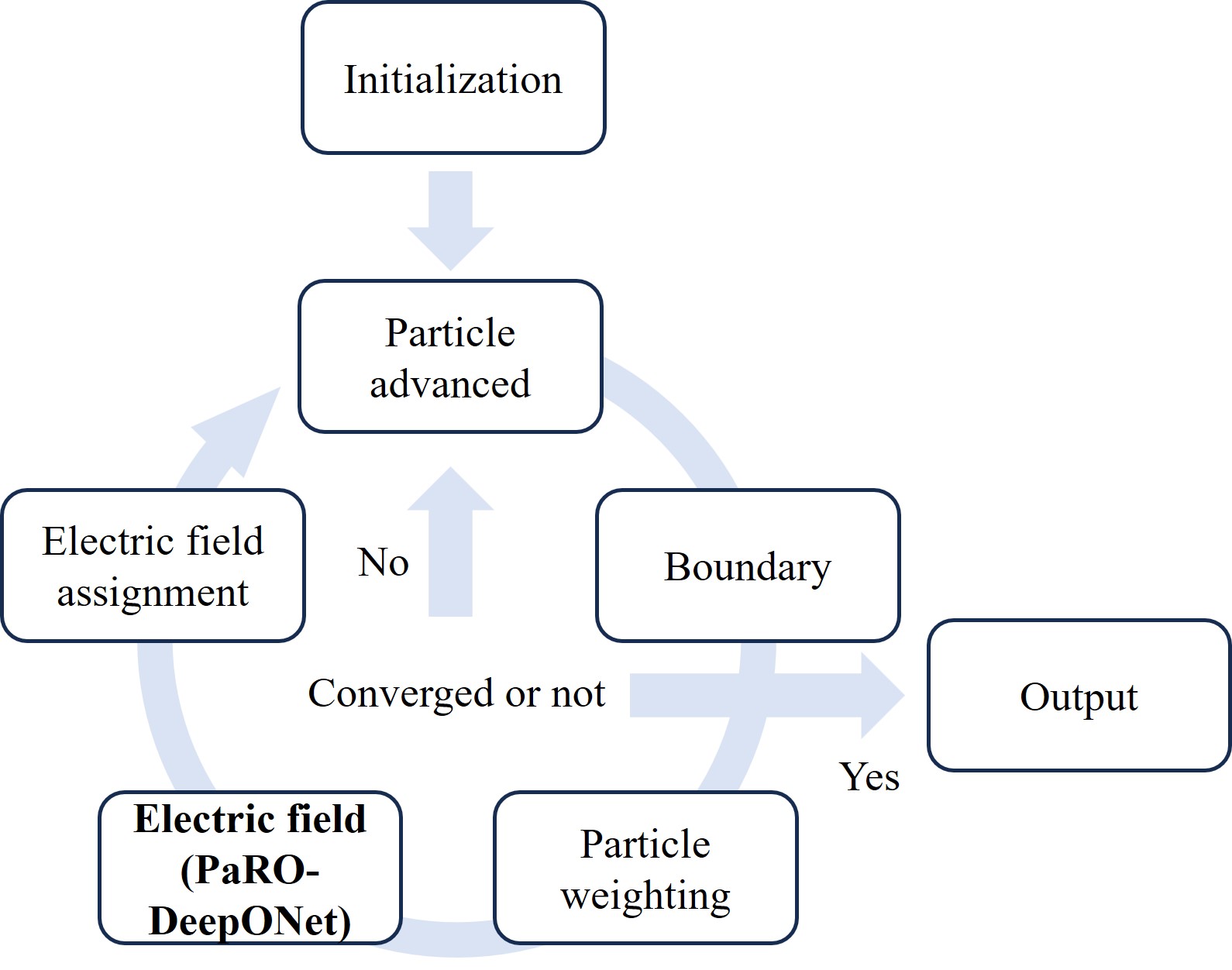}
	\caption{Electrostatic PIC simulation framework with PaRO-DeepONet as a surrogate Poisson solver.}
	\label{fig:fig2}
\end{figure}

\paragraph{}
The objective of PaRO-DeepONet in this work is not merely to learn a mapping from charge density distributions to their corresponding electrostatic potentials. Rather, the model is designed to be integrated into a complete electrostatic Particle-In-Cell (PIC) simulation framework, acting as a surrogate Poisson solver that maintains fidelity to the classical numerical results while offering improved computational efficiency. The overall architecture is illustrated in Figure \ref{fig:fig2}.
\paragraph{}
To this end, we develop an in-house electrostatic PIC code (ESPIC). This code follows the standard PIC workflow: it interpolates particle charges onto a uniform grid, solves the Poisson equation to compute the electrostatic field, and then interpolates the field back to the particle positions to update their motion. The key innovation in our approach lies in replacing the traditional Poisson solver with PaRO-DeepONet. This enables rapid and data-driven solutions to the electrostatic field without requiring the direct solution of the Poisson equation at every time step.

\paragraph{}
To ensure consistency and accuracy, the training dataset for PaRO-DeepONet is generated within the ESPIC framework, where the Poisson equation is solved via classical FDM-based solver. This ensures that the learned surrogate model is well-aligned with the discretization schemes, boundary conditions, and particle-grid interactions inherent in the PIC simulations.

\paragraph{}
Further implementation details of ESPIC, including particle injection, grid discretization, and boundary condition treatments, are presented in Section 3.

\label{sec:sec3}
\section{Numerical experiments}

\paragraph{}
To evaluate the performance of PaRO-DeepONet, we present comprehensive experiments involving four distinct ESPIC models. These cases vary in geometry, electron distribution, and boundary conditions to validate the model under diverse physical configurations. The training datasets used to construct PaRO-DeepONet and its performance as a replacement for the conventional FDM solver are discussed in detail. The tolerance level of FDM solver is set to $10^{-16}$. All simulations are implemented using the open-source machine learning framework PyTorch \cite{31}, and the neural networks are trained using the stochastic optimization algorithm Adam \cite{32} with either a constant or decaying learning rates.

\label{sec:sec3.1}
\subsection{Diffusion of 2D electrostatic sphere}

\paragraph{}
This test case simulates the evolution of particle dynamics in a 2D electrostatic field induced by a centrally located spherical charge distribution. The computational domain is defined as $x = y \in [-0.5, 0.5] $ m with a spatial resolution of $n_x$=$n_y$=64. Perfect electric conductor (PEC) boundary conditions are imposed on the fields, while absorbing conditions are applied to the particles. Particle injection follows a uniform distribution, with 6 particles initialized per spatial dimension. The radius of the electrostatic sphere is treated as the variable initial condition. The simulation runs until the number of particles within the grid decreases to 10$\%$ of the initial count. The time step is set to $\Delta t=1\times 10^{-5}$ s, and each simulation runs for approximately 200 time steps.

\paragraph{}
The initial radii used for training are [0.1, 0.11, 0.12, 0.13, 0.15, 0.16, 0.17, 0.18], while the test simulation is performed using a radius of 0.14. To accelerate training, all 64×64 grid data are downsampled to 16$\times$16. The same downsampling strategy is consistently applied to all subsequent experiments.

\begin{figure}
	\centering
	\includegraphics[width=15.5cm]{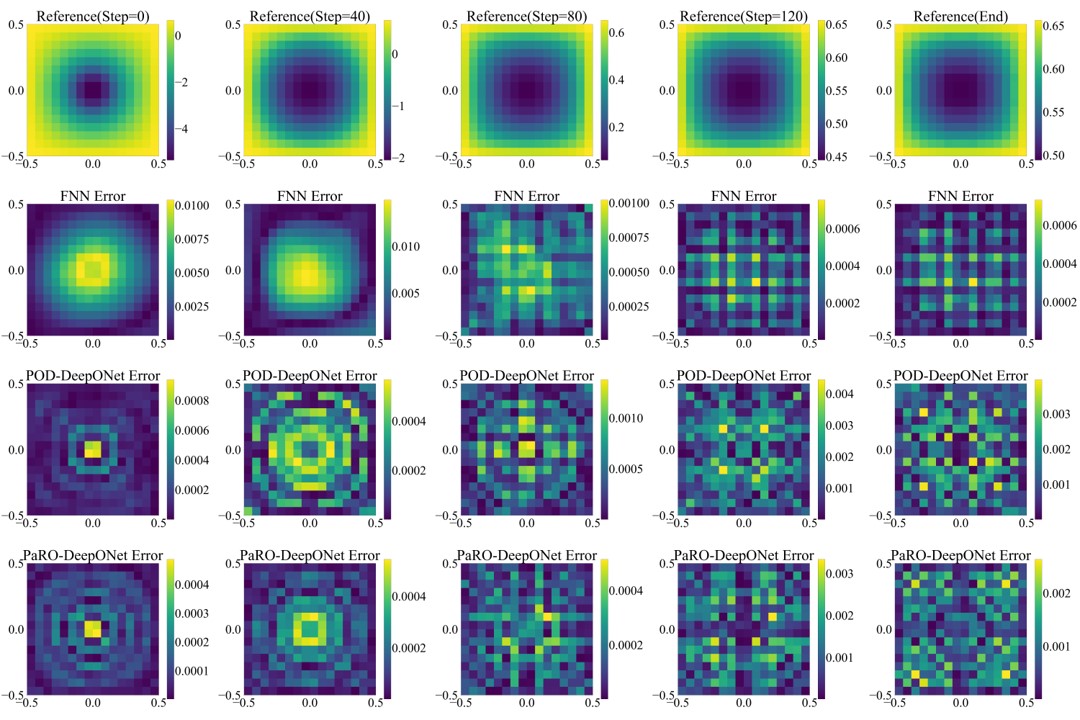}
	\caption{Diffusion of 2D electrostatic sphere:Relative L$^2$ error distribution of potential predicted by FNN, POD-DeepONet, and PaRO-DeepONet across 5 snapshots with reference solution by FDM.}
	\label{fig:fig3}
\end{figure}

\paragraph{}
To demonstrate the performance of PaRO-DeepONet, we compare it against traditional DeepONet with fully connected branch network (hereinafter referred to as FNN) and POD-DeepONet. As illustrated in Figure \ref{fig:fig3}, FNN directly inputs the 4096-point charge density distribution into the branch network, whereas POD-DeepONet and PaRO-DeepONet reduce the input dimension to 311 and 733, respectively. The branch network architecture for all models is defined as [input dim, 128, 128, 128, 128, 128, 64], and the trunk network as [2, 128, 128, 128, 128, 128, 64].

\paragraph{}
The relative L$^2$ errors of FNN, POD-DeepONet, and PaRO-DeepONet are 0.032, 0.0102, and 0.0090, respectively. As shown in Figure \ref{fig:fig3}, FNN fails to effectively learn the mapping from charge density distribution $\rho$ to potential $\varphi$, while POD-DeepONet improves accuracy by compressing input dimensions but still struggles with capturing localized features. PaRO-DeepONet further enhances performance by embedding particle information, leading to significantly smoother error distributions and more accurate predictions, especially near the center of the charge distribution.

\begin{figure}
	\centering
	\includegraphics[width=15.5cm]{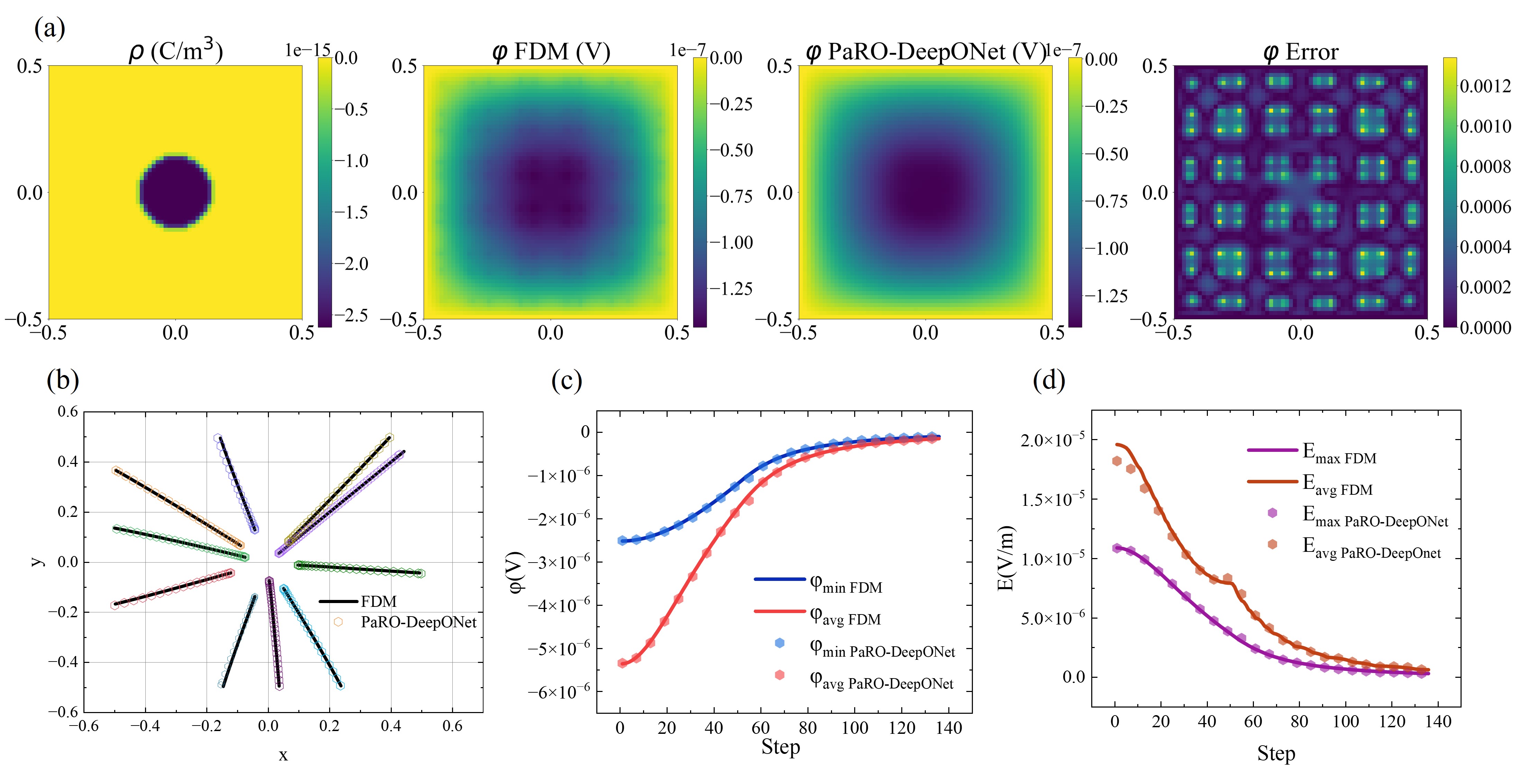}
	\caption{Diffusion of 2D electrostatic sphere: (a) Initial charge density distribution, grouth-truth solution of potential by FDM at the final time step, PaRO-DeepONet prediction, and relative L$^2$ error distribution. (b) Comparison of ten particle trajectories. (c) Evolution of minimum and mean values of electric potential at x=0. (d) Evolution of maximum and mean values of electric field at x=0.}
	\label{fig:fig4}
\end{figure}

\paragraph{}
After training, PaRO-DeepONet is integrated into the ESPIC framework, replacing the FDM solver. The final simulation results are shown in Figure \ref{fig:fig4}. The total simulation time of ESPIC using FDM and PaRO-DeepONet are 52.78 s and 1.66 s, respectively. Notably, the Poisson solver time is reduced from 51.69 s (FDM) to 0.18 s (PaRO-DeepONet). The relative L$^2$ error of PaRO-DeepONet in the final step is 0.0237.

\paragraph{}
Figure \ref{fig:fig4} presents the final simulation results under the same test conditions. As observed in Fig. \ref{fig:fig4}(a), PaRO-DeepONet reproduces the potential distribution predicted by FDM with high fidelity. Particle trajectory comparisons in Fig. \ref{fig:fig4}(b) show minimal deviation between the two solvers. Figures \ref{fig:fig4}(c) and \ref{fig:fig4}(d) further confirm that the electric potential and field along the geometric center line remain consistent throughout the evolution, indicating no long-term error accumulation.

\paragraph{}
The interesting observation is that PaRO-DeepONet does not yield a discontinuous potential distribution despite the discretization of charge. As the simulation progresses, particles are pushed apart by electrostatic forces, resulting in increasingly sparse and isolated charge deposition. Under such circumstances, one would expect the potential field to become equally discontinuous, as reflected in the FDM results. However, the potential predicted by PaRO-DeepONet remains remarkably smooth, with discrepancies mainly occurring at isolated grid points. This may be attributed to the inherent expressiveness of DeepONet, which is capable of adapting to data with varying resolutions \cite{33}. Its ability of producing high-resolution outputs from low-resolution inputs allows it to maintain continuity in the predicted fields. This behavior is especially advantageous in PIC simulations, where discretization-induced errors are notoriously difficult to eliminate. Traditional approaches to mitigate this problem often require more refined deposition techniques or increased particle counts—both of which substantially increase computational cost. Therefore, PaRO-DeepONet offers a promising solution by naturally suppressing discretization artifacts without incurring additional computational burden.

\label{sec:sec3.2}
\subsection{Diffusion of electrons in L-shaped domain}

\paragraph{}
To further evaluate the generalization capability of PaRO-DeepONet, this case considers a 2D L-shaped domain defined as $\Omega $=(0,1)$^2$$\setminus$ [0.5,1)$^2$, discretized with a spatial resolution of $n_x$ = $n_y$ = 64. PEC boundary conditions are applied on the field boundaries at $x$ =1 and $y$ =1 m, each held at 1$\times$10$^{-6}$ V. Absorbing boundaries are used for particles along the same edges. The remaining field boundaries are set as insulating, and the remaining particle boundaries are reflecting. The initial particle distribution is randomly generated via Gaussian random fields (GRF), assigning 0$-$2 particles per cell per dimension. A training dataset of 1000 charge distributions is generated using a fixed random seed, and a test simulation is run using a previously unseen distribution. The simulation is terminated when the particle count drops to 20$\%$ of its initial value, with a time step of $\Delta $t =5$\times$10$^{-5}$ s and approximately 55 total time steps. The hidden layer architecture for both branch and trunk networks is set as [256, 128, 128, 128]. The solution time of ESPIC with and without PaRO-DeepONet is 14.86 s and 8.75 s respectively, and the time for Poisson solution is 13.73 s and 0.065254 s respectively. The relative L$^2$ error of PaRO-DeepONet in the final step is 0.0323.

\begin{figure}
	\centering
	\includegraphics[width=15.5cm]{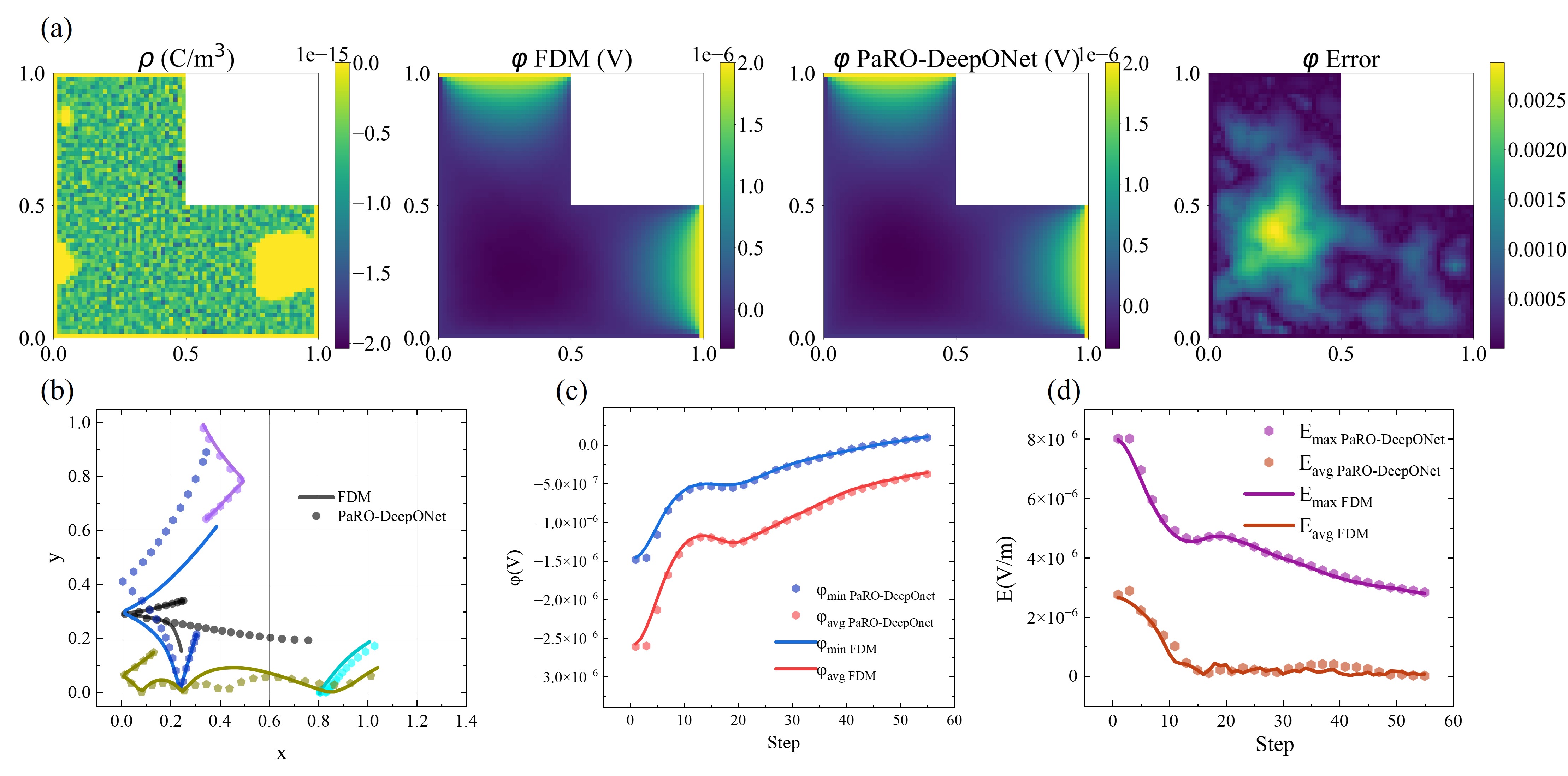}
	\caption{Diffusion of electrons in L-shaped domain: (a) Initial charge density distribution, grouth-truth solution of potential by FDM at the final time step, PaRO-DeepONet prediction, and  relative L$^2$ error distribution. (b) Comparison of five particle trajectories. (c) Evolution of minimum and mean values of electric potential at x = 0.25 (d) Evolution of maximum and mean values of electric field at x = 0.25.}
	\label{fig:fig5}
\end{figure}

\paragraph{}
Although the performance in this case is slightly inferior to Case I in Section 3.1, PaRO-DeepONet still achieves a reasonable approximation of the potential field. The errors may stem from two factors: (1) the stochastic nature of the charge distributions may pose challenges to the POD-based dimensionality reduction, and (2) the presence of complex reflection boundary conditions may amplify the propagation of electric field error during the simulation. Nevertheless, the model still significantly accelerates the overall computation while maintaining physically acceptable accuracy, thereby demonstrating its robustness when applied to irregular geometries and stochastic input conditions.

\label{sec:sec3.3}
\subsection{Electrons extraction}

\paragraph{}
This case investigates a more realistic and complex geometry based on electron extraction systems in fusion neutral beam sources \cite{34}. The domain features an extraction hole through which only electrons are allowed to propagate. Initially, electrons are confined within a rectangular region near $x = 1\sim 1.7\times 10^{- 3}$ m, with 3 particles per dimension. Periodic boundary conditions are applied to both particles and fields at $y = 0$ and $y = y_{max}$. A fixed potential of 5$\times$10$^{-4}$ V is applied at $x = x_{max}$, and the potential at $x = 0$ is treated as the variable initial condition.

\paragraph{}
The training dataset consists of potential values in the set [6, 7, 8, 9, 11, 12e, 13, 14, 15]$\times$10$^{-5}$ V, with the test case conducted at 1$\times$10$^{-5}$ V. The potential value is appended to the modal vector as an additional input. The simulation is terminated once 90$\%$ of the particles have exited the domain, with a time step of $\Delta $t = 1$\times$10$^{-8}$ s and approximately 70 time steps in total. Both the branch and trunk networks adopt the architecture as [128, 128, 128, 128], and a hard-coded symmetry constraint is applied to the output of the PaRO-DeepONet:

\begin{equation}
	\label{equ:equ8}
\widehat {\varphi '} = \frac{{\widehat \varphi  + mirror\left( {\widehat \varphi } \right)}}{2}
\end{equation}

where $\widehat{\varphi }$ is the original output of the PaRO-DeepONet,  $mirror\left( {\cdot } \right)$ denotes the reflection operator along the $x$-axis, and $\widehat{\varphi'}$  is the final output after symmetry correction. 

\paragraph{}
The total simulation time using the original ESPIC solver is 339.02 s, which is reduced to 70.62 s when using PaRO-DeepONet. The Poisson solver time is reduced from 270.34 s to 0.9 s. The relative L$^2$ error of the PaRO-DeepONet prediction at the final time step is 0.0157.

\begin{figure}
	\centering
	\includegraphics[width=15.5cm]{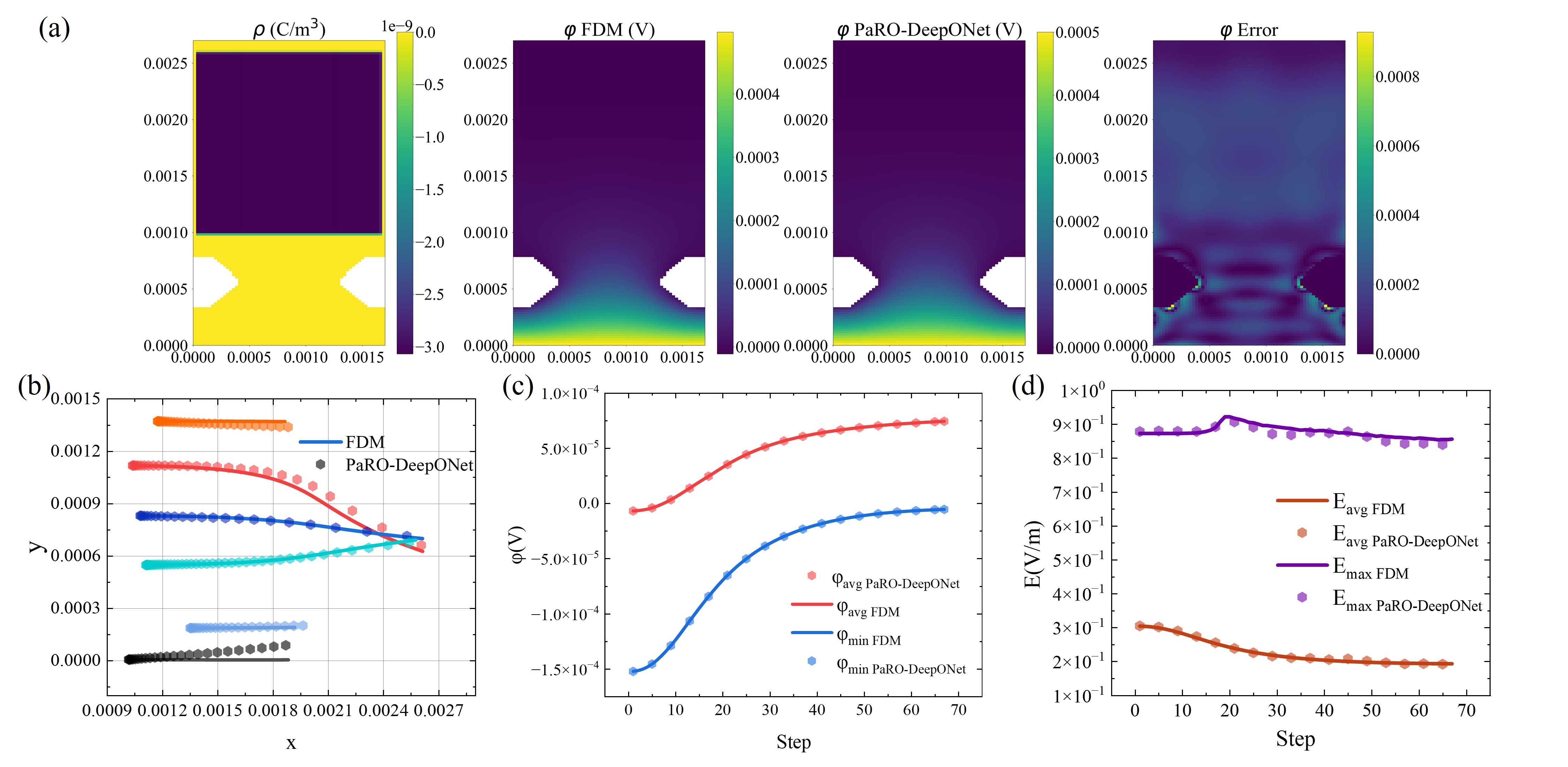}
	\caption{Electrons extraction: (a) Initial charge density distribution, grouth-truth solution of potential by FDM at the final time step, PaRO-DeepONet prediction, and relative L$^2$ error distribution. (b) Comparison of six particle trajectories. (c) Evolution of minimum and mean values of electric potential at y = 0.00075. (d) Evolution of maximum and mean values of electric field at y = 0.00075.}
	\label{fig:fig6}
\end{figure}

\paragraph{}
As shown in Figure \ref{fig:fig6}, PaRO-DeepONet achieves high-fidelity predictions in this geometrically and physically complex setting. Only minor deviations appear near the y-boundaries, likely due to edge effects under periodic conditions. Nonetheless, this case demonstrates the model’s adaptability to sophisticated geometries and its compatibility with symmetry-enforcing constraints, highlighting its strong potential for real-world applications in electrostatic field modeling.

\label{sec:sec3.4}
\subsection{Diffusion of 3D electrostatic sphere}

\paragraph{}
When employing PIC methods to study complex plasma phenomena, three-dimensional (3D) modeling is often essential to accurately capture spatial characteristics. However, 3D simulations result in an exponential increase in grid points, which substantially raises computational costs and makes iterative convergence more challenging. In this case, we extend the case in Section 3.1 to three dimensions in order to evaluate the generalization performance of PaRO-DeepONet in high-dimensional scenarios with sparse charge distributions.

\paragraph{}
The computational domain is defined as $x =y=z \in [-0.5, 0.5]$ m, with spatial discretization steps $n_x=n_y=n_z=64$. The boundary conditions are consistent with those in the case of Section 3.1. Particles are injected non-uniformly, with two particles per cell per dimension. The radius of the electrostatic sphere serves as the initial condition parameter. Under a fixed time step $\Delta $t =1$\times$10$^{-6}$ s, each simulation runs for approximately 200 steps until the particle count drops to 10$\%$ of its initial value.

\paragraph{}
To construct the training dataset, simulations are performed at initial radii $R$ = [0.3, 0.304, 0.306, 0.308, 0.310, 0.312, 0.314, 0.316, 0.320, 0.322, 0.324, 0.326, 0.328, 0.330, 0.332, 0.334, 0.336] m. The test case is conducted with $R$ = 0.318 m. The radius step is finer than in the case of Section 3.1, due to the fact that even minimal changes in radius in 3D space lead to significant particle count variations—for instance, increasing from 235712 particles at $R$ = 0.3 m to 281664 at$R$ = 0.318 m.

\paragraph{}
Both the trunk and branch networks adopt the architecture configured as [128, 128, 128, 128]. Due to the low magnitude of the electric field in the Z-direction in 3D simulations—often close to the error scale of PaRO-DeepONet—this can cause particle motion to deviate from physical expectations. To address this, we introduce gradient regularization to suppress overfitting and ensure a more uniform error distribution across the spatial domain. The regularized output is given by:

\begin{equation}
		\label{equ:equ9}
L = \frac{{{{\left\| {\widehat{\varphi } - \varphi } \right\|}_2}}}{{{{\left\| \varphi  \right\|}_2}}} + \lambda \sum\limits_\theta  {{{\left\| {{\nabla _\theta }{L_{data}}} \right\|}_2}} 
\end{equation}

where L is the loss function, $\widehat{\varphi }$  and $\varphi$  are the predicted potential and the ground truth, respectively, $ \nabla _\theta{L_{data}}$ is the gradient of the data loss with respect to the model parameters $\theta $, and $\lambda $ is a weighting coefficient that controls the strength of the gradient regularization term. In this case, $\lambda $ = 0.1.

\begin{figure}
	\centering
	\includegraphics[width=15.5cm]{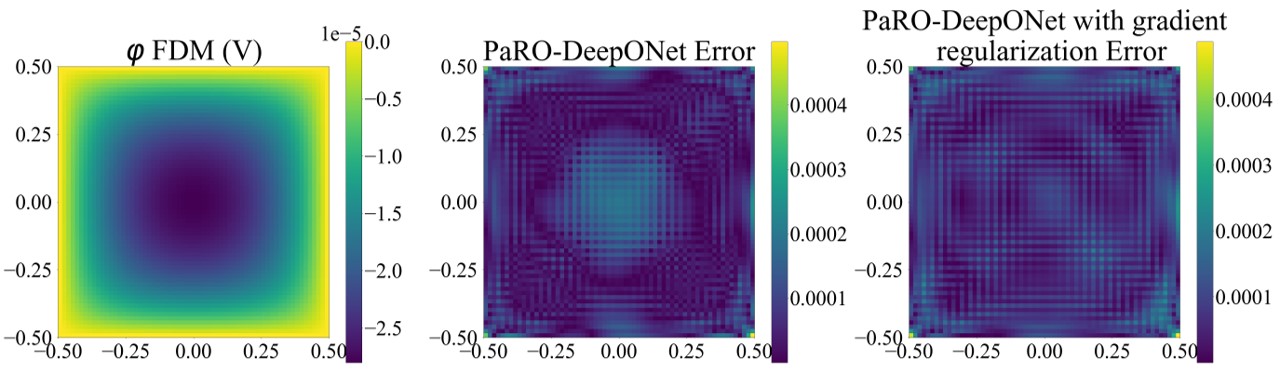}
	\caption{Diffusion of the 3D electrostatic sphere:Relative L$^2$ error distribution of PaRO-DeepONet predictions with and without regularization on the z = 0 slice at the final time step, compared against the reference solution obtained by FDM under identical initial conditions.}
	\label{fig:fig7}
\end{figure}

\paragraph{}
When error distribution is spatially uneven, especially along weak field directions such as Z, the resulting electric field becomes spatially asymmetric, leading to inaccurate particle trajectories. However, as shown in Figure \ref{fig:fig7}, the use of gradient regularization significantly reduces spatial error variance, helping restore isotropy in the electrostatic force field and ensuring particles are influenced symmetrically in all directions.

\begin{figure}
	\centering
	\includegraphics[width=12.5cm]{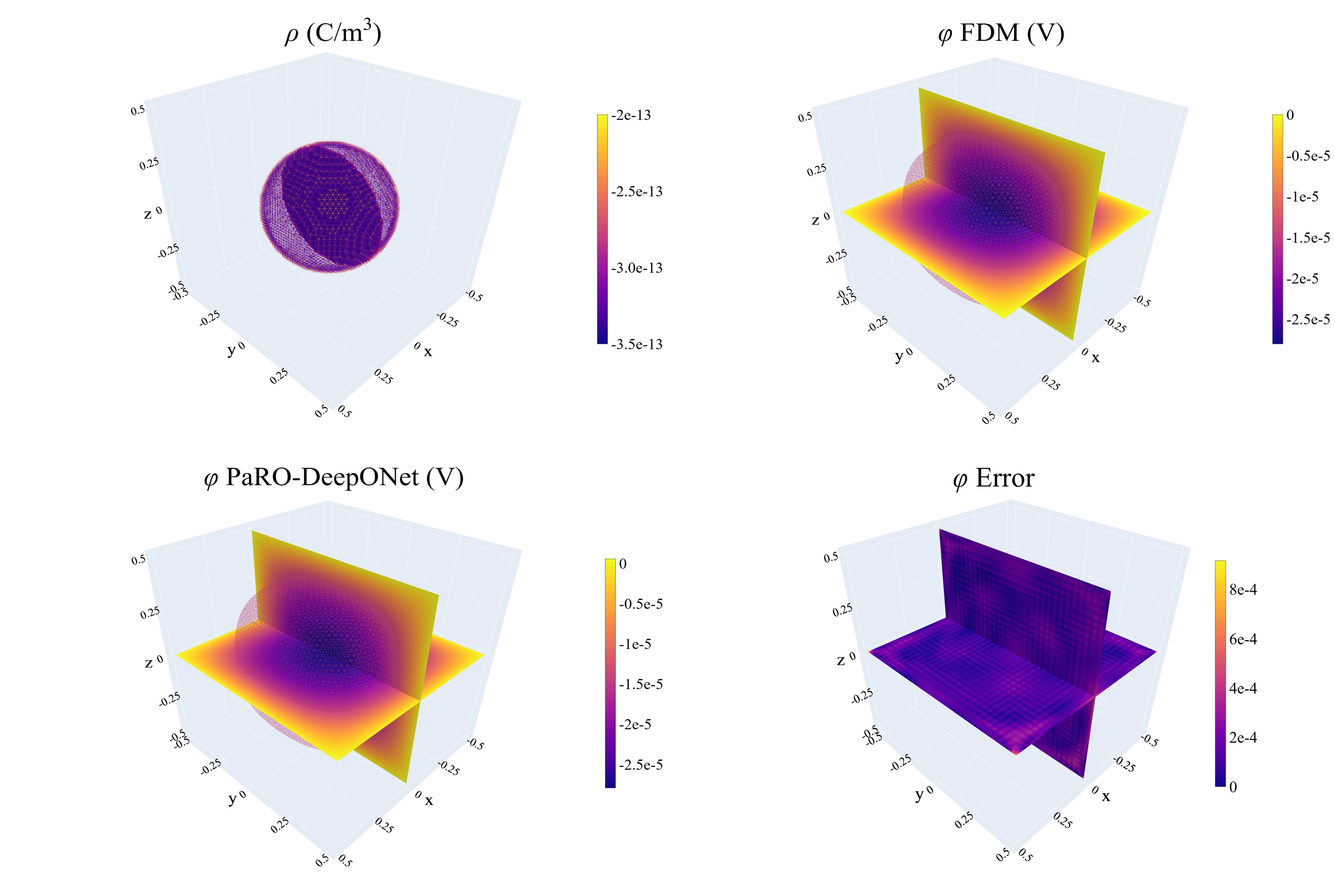}
	\caption{Diffusion of 3D electrostatic sphere: Initial charge density distribution, grouth-truth solution of potential by FDM at the final time step, PaRO-DeepONet prediction, and relative L$^2$ error distribution.}
	\label{fig:fig8}
\end{figure}

\begin{figure}
	\centering
	\includegraphics[width=12.5cm]{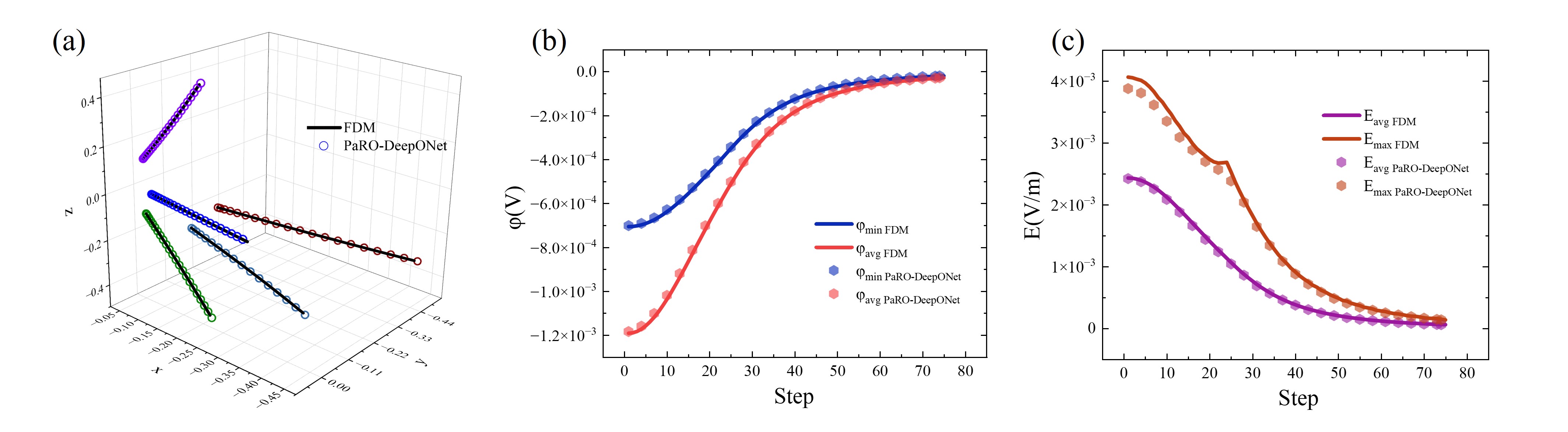}
	\caption{Diffusion of 3D electrostatic sphere: (a) Comparison of five particle trajectories. (b) Evolution of minimum and mean values of electric potential at x = 0 and y = 0. (c) Evolution of maximum and mean values of electric field at x = 0 and y = 0.}
	\label{fig:fig9}
\end{figure}

\paragraph{}
The runtime of the ESPIC simulation without and with PaRO-DeepONet is 270.2 s and 74.17 s, respectively. Specifically, the Poisson solver time is reduced from 204.03 s to 27.14 s. The relative L$^2$ error of the PaRO-DeepONet prediction at the final step is 0.0082.

\paragraph{}
Figures \ref{fig:fig8} and \ref{fig:fig9} present a comprehensive overview of the simulation results. Figure \ref{fig:fig8} compares the initial charge density, the final FDM solution of potential, the PaRO-DeepONet prediction, and the error distribution. Despite the increased in dimensionality and data sparsity, PaRO-DeepONet consistently delivers high-fidelity predictions and stable particle trajectories, performing comparably to the lower-dimensional case in Section 3.1. This confirms the potential of PaRO-DeepONet to handle 3D sparse systems, which are traditionally challenging for both conventional solvers and neural operators due to their low signal-to-noise ratio and sensitivity to spatial symmetry.

\label{sec:sec4}
\section{CONCLUSIONS}

\paragraph{}
In this work, we proposed PaRO-DeepONet, a particle-informed reduced-order surrogate model that integrate Proper Orthogonal Decomposition (POD) with Deep Operator Networks (DeepONet) to accelerate the solution of Poisson equations in electrostatic Particle-in-Cell (PIC) simulations. By directly learning the mapping from charge density to potential fields, PaRO-DeepONet bypasses traditional iterative solvers, significantly reducing computational costs while maintaining high accuracy.

\paragraph{}
The effectiveness of the proposed framework was validated on four representative test cases of increasing complexity: 2D and 3D electrostatic sphere expansion, diffusion of electrons in L-shaped domain, and electrons extraction. In all cases, PaRO-DeepONet exhibited strong predictive capability, achieving relative L$^2$ errors below 3.5$\%$ and reducing Poisson solver runtime by up to 99.6$\%$. Notably, in the 2D case, the model successfully reconstructed smooth potential fields from highly discrete charge distributions—an important feature for practical PIC applications where particle sparsity often leads to noise-induced error in traditional solvers. In the 3D case, a gradient-based regularization strategy was introduced to mitigate asymmetry in weak field regions, further enhancing prediction stability and physical consistency.

\paragraph{}
These results demonstrate that PaRO-DeepONet not only provides a scalable and accurate alternative to classical Poisson solvers but also offers a promising direction for addressing longstanding challenges in kinetic plasma modeling, such as sparse feature capture, grid resolution limitations, and computational bottlenecks in high-dimensional domains.

\paragraph{}
In future work, we plan to extend this surrogate framework to dynamic charge-potential coupling, adaptive mesh environments, and multi-physics integration, thereby enabling fully end-to-end learning-based PIC pipelines for large-scale plasma simulations.

\section*{Acknowledgments}
\label{sec:acknowledgments}
\paragraph{}
This work was supported in part by the National Natural Science Foundation of China (92470102) and the Natural Science Foundation of Jiangsu Province (BK20231427).

\end{document}